\documentstyle[epsfig,12pt]{article}

\newcommand{\nc}{\newcommand}
\nc{\beq}{\begin{equation}} \nc{\eeq}{\end{equation}}
\nc{\beqa}{\begin{eqnarray}} \nc{\eeqa}{\end{eqnarray}}
\begin{document}

\title{{\large {\bf Lightsheets, Branes and the Covariant Entropy Bound}}}
\author{Deog Ki Hong\thanks{dkhong@pusan.ac.kr} \\
Department of Physics\\
Pusan National University, Pusan 609-735, Korea \\
\\
Stephen D.H.~Hsu\thanks{
hsu@duende.uoregon.edu} \\
Department of Physics \\
University of Oregon, Eugene OR 97403-5203 \\}
\date{\today}
\maketitle

\begin{abstract}
We investigate the geometry of lightsheets comprised of null
geodesics near a brane. Null geodesics which begin parallel to a
brane a distance d away are typically gravitationally bound to the
brane, so that the maximum distance from the geodesic to the brane
never exceeds d. The geometry of resulting lightsheets is similar
to that of the brane if one coarse grains over distances of order
d. We discuss the implications for the covariant entropy bound
applied to brane worlds.
\end{abstract}

\newpage

It is possible that the maximum information content of a spacetime
region is related to its surface
area~\cite{'tHooft:gx,Susskind:1994vu,Bousso:2002ju}. The idea has
its origins in the proposal of Bekenstein that the area of a black
hole is proportional to its entropy \cite{Bekenstein:ur}, and that
black holes obey a generalized second law of thermodynamics (GSL)
\cite{Bekenstein:ax}. A covariant generalization of these ideas
\cite{Fischler:1998st,Bousso:1999xy} has passed a number of
theoretical tests, and suggests a deep relationship between
geometry and information which arises due to quantum gravity. This
covariant entropy bound (henceforth, the covariant bound) can be
stated as follows:

{\it Let $A(B)$ be the area of an arbitrary $D-2$ dimensional
spatial surface $B$, which need not be closed. A $D-1$ dimensional
hypersurface $L(B)$ is the light-sheet of $B$ if $L(B)$ is
generated by light rays extending orthogonally from $B$, which
have non-positive expansion everywhere on $L(B)$. Let $S(L)$ be
the total entropy of matter which intersects $L(B)$. Then $S(L)
\leq {1 \over 4} A(B)$.}

For simple cases, such as a suitable closed spacelike surface
surrounding a weakly gravitating system, the covariant bound
reduces to the usual area bound.

\begin{figure}
\epsfxsize=1in \centerline{\epsffile{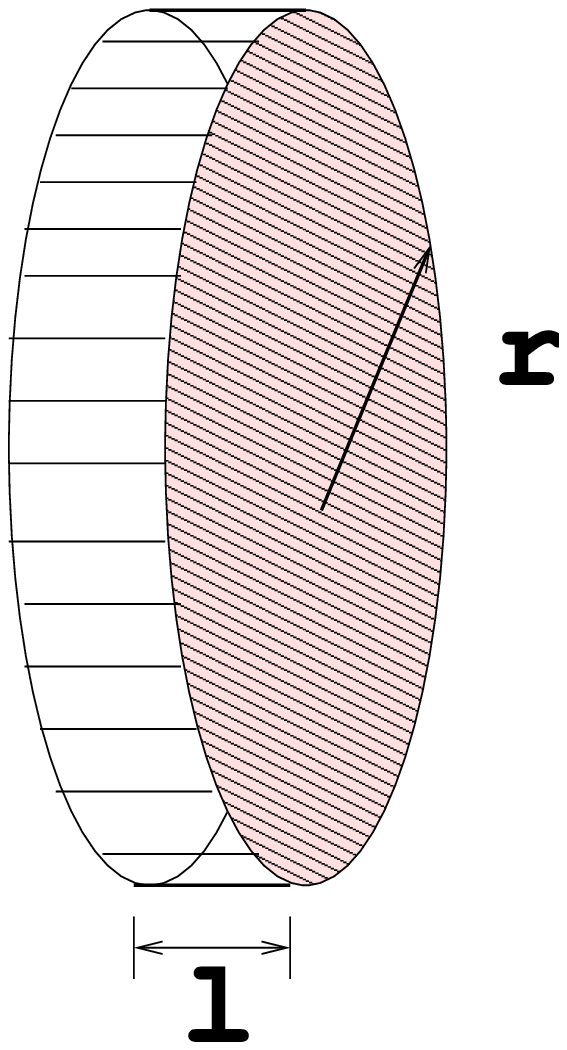}}
%
%\centerline{\includegraphics[scale=0.5]{hat.eps}}%
\caption{$D-2$ dimensional surface B (the unshaded region of
thickness $l$) in the brane scenario.}
 \label{fig1}
\end{figure}

In \cite{HH}, the covariant bound was shown to be violated in
brane world scenarios \cite{ADD,RS} in which the fundamental scale
of quantum gravity is $M_* \sim {\rm TeV}$. Consider a spacelike
region V of extent $r$ on the 3-brane and thickness $l$ in the
orthogonal extra dimensions (see Fig.~{\ref{fig1}}). The boundary
of V consists of components whose surface areas scale as $r^3 \,
l^{(D-5)}$ and $r^2 \, l^{(D-4)}$. The first surface component is
obtained by setting the extra-dimensional coordinates at their
extreme (boundary) values and allowing the coordinates $\{ x_{1-3}
\}$ to vary throughout the intersection of V with the 3-brane.
(This is shown as the shaded region in Fig.~{\ref{fig1}}.) The
second is obtained by setting $\{ x_{1-3} \}$ at their extreme
values (i.e., the boundary on the 3-brane) and letting the
extra-dimensional coordinates to vary over a range of size $l$.
(This is indicated by the unshaded, but lined, region in the
figure.)

The surface $B$ in the covariant bound is taken as the {\it
second} part of the boundary of $V$, the one whose area scales as
$r^2 \, l^{(D-4)}$ (note that $B$ need not be closed). In Fig.
{\ref{fig1}} this appears as the unshaded portion of the
cylindrical surface. Let $V$ have the same shape as the brane,
with thickness $l$ of order $M_*^{-1}$ (the minimum thickness
possible; the same as that of the brane), so that its surface area
is of order $r^2$ in $M_*$ units. The light sheet $L(B)$ is
comprised of null geodesics emanating orthogonally from $B$. These
geodesics intersect all of the ordinary matter in $V$, so the
entropy $S(L)$ is simply that of the ordinary matter in $V$. In
\cite{HH} it was shown that $S(L)$ can exceed $A(B)$ in $M_*$
units for systems such as a supernova core or the early universe.

\begin{figure}
\epsfxsize=3in \centerline{\epsffile{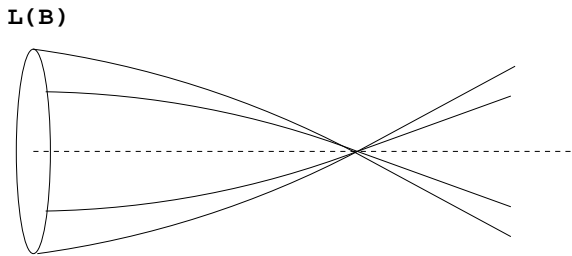}}
%
%\centerline{\includegraphics[scale=0.5]{hat.eps}}%
\caption{Null geodesics focused into caustic.}
\label{fig2}
\end{figure}

Taking $B$ to be the same thickness of the brane avoids the
question of whether the gravitational pull of the brane in the
extra dimension focuses the rays of $L(B)$ to a caustic before
they reach the center of the fiducial volume. The condition of
non-expansion used in defining $L(B)$ would cause it to terminate
at a focal point, and much of the matter would never intersect
$L(B)$. (See Fig.~{\ref{fig2}})
Because $l_* \sim M_*^{-1}$ is the fundamental length scale of
quantum gravity and also the thickness of the brane, we do not
consider any focusing of $L(B)$. (There is likely no meaning to
distances less than $l_*$ \cite{minlength}.) However, it remains
to be seen whether this $L(B)$ can be obtained as a smooth limit
of lightsheets $L(B')$ resulting from surfaces $B'$ with larger
extent in the extra dimensions. Otherwise, one might consider the
construction used in \cite{HH} to be a degenerate
limit\footnote{S. Hsu thanks S. Giddings for emphasizing this
point.}. We address this issue below, and show that a lightsheet
$L(B)$ with the same geometry as the brane can be obtained as the
smooth limit of a family of lightsheets $L(B')$, if a slightly
modified (coarse grained) definition of lightsheets is adopted.
The gravitational binding of null geodesics to a nearby brane is a
key component of this analysis.

In RS geometry \cite{RS} the metric is given
as~\cite{Giddings:2000mu}
\begin{eqnarray}
\label{RS}
ds^2=dy^2+e^{-2|y|/R}\eta_{\mu\nu}dx^{\mu}dx^{\nu}.
\end{eqnarray}
The equations for geodesics emanating from orthogonally from $B$,
and parallel to the TeV brane located at the origin, $y=0$, are
given as, for $y\ge0$,
\begin{eqnarray}
{d^2y\over ds^2} ~+~ {2\over R}e^{-2y/R}\left[\left({dt\over
ds}\right)^2- \left({d\vec x\over ds}\right)^2\right] ~&=&~ 0
\label{g1}
\\ \
{d^2t\over ds^2} ~+~{dt\over ds}{dy\over ds } ~&=&~ 0\\
{d^2\vec x\over ds^2} ~-~ {d\vec x\over ds}{dy\over ds } ~&=&~ 0
\label{g3} ~.
\end{eqnarray}
For this special geometry, solutions with $dy/ds = 0$ always exist
if the velocity vector $u^a = (dt/ds, d\vec{x}/dt)$ is null (i.e.,
the difference in brackets in equation (\ref{g1}) vanishes).
Therefore, it is possible to have light rays which travel parallel
to the brane without being focused. If the rays remain parallel to
the brane, the lightsheet $L(B)$ is easily seen to result from the
smooth limit of lightsheets $L(B')$, which are similar to $L(B)$
but with larger extent in the extra dimensions.

In general the metric on the brane is not simply $\eta_{\mu \nu}$,
i.e., if there is matter or energy density on the brane. In the
examples considered in \cite{HH}, the metric on the brane is
either the Robertson-Walker metric of the early universe, or that
of a supernova interior, and may have $t$ or $\vec{x}$ dependence.
In such cases we expect that geodesics may be bent toward the
brane by gravitational attraction. Consider a small perturbation
to the metric (\ref{RS}), and suppose that it forces the solution
to the geodesic equations to have non-zero $dy/ds \neq 0$. Since
the deviation from (\ref{RS}) is assumed to be a perturbation, we
deduce the leading order behavior as follows. The general solution
to (\ref{g1}-{\ref{g3}) for arbitrary initial conditions satisfies
\begin{eqnarray}
{dt \over ds} &=&C_0 e^{-y}\quad  \\
\frac{dx_i}{ds} &=& C_ie^{y}\\
{1\over2}\left({dy\over ds}\right)^2 &=& E - V(y)~,
\end{eqnarray}
where $C_0,C_i$ and $E$ are constants of integration, and
\begin{equation}
V(y)={-2C_0^2\over 2R+2}e^{-(2/R+2)y}-{2C_i^2\over 2R-2}e^{(2-2/R)y}\,.
\end{equation}
We see that in general the geodesics hit the origin $y=0$ exponentially quickly.

\begin{figure}
\epsfxsize=3.5in \centerline{\epsffile{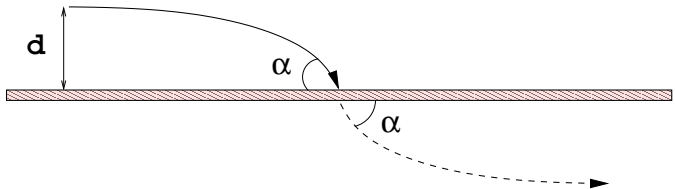}}
%
%\centerline{\includegraphics[scale=0.5]{hat.eps}}%
\caption{Focusing of light rays due to the gravitational force of
the brane.}
 \label{fig3}
\end{figure}

However, if a light ray, starting parallel to a brane $M$, hits
$M$ after some time, it will be forever bound to $M$. The proof is
as follows (see Fig.~{\ref{fig3}}). Let $M$ be infinite and
uniform. Assume the $D$-dimensional universe is $Z_2$ symmetric
under reflections through $M$. Let a light ray start parallel to
$M$ at height $d$ and intersect $M$ from above at point $P$ with
angle $\alpha$. Consider the time reversed trajectory: it shows
that a ray leaving $M$ with angle $\alpha$ will eventually end up
parallel to $M$ at height $d$. Now, when the original ray crosses
through $M$ from above at point $P$ it then leaves $M$ from below
at angle $\alpha$. By symmetry (using $Z_2$ reflections through
$M$ itself, and through a hyperplane orthogonal to $M$ which
passes through $P$), this ray will end up parallel to the brane at
a height $-d$. This argument, when repeated, implies the ray is
bound forever to $M$ and never more than distance $d$ away from
$M$. Although we used translation invariance in the above argument
(to justify the $Z_2$ reflection through the hyperplane orthogonal
to $M$ at $P$), we expect corrections due to small deviations from
translation invariance to be small. Indeed, the question of
whether a light ray is bound to the brane (i.e., whether it can
escape to infinity) is ultimately an energetic one, and hence not
sensitive to small rearrangements of the energy on the brane.

The binding of null geodesics to a brane seems outside the range
of behaviors imagined for lightsheets in the formulation of the
covariant bound \cite{Bousso:2002ju}. In the original formulation,
an important objective was to give a criteria for terminating a
lightsheet, since an infinite lightsheet might intersect an
infinite amount of entropy and render the bound problematic. It
seems to have been implicitly assumed that focusing of light rays,
or the formation of a caustic, would inevitably lead to subsequent
expansion and divergence of the lightsheet area (as depicted in
Fig.~{\ref{fig2}}). The argument above shows that focusing does
not necessarily imply divergent expansion.

Given the semiclassical spirit of the covariant entropy bound, it
seems reasonable to use a {\it coarse grained} definition of
lightsheets $L(B)$. In particular, in deciding where to terminate
a lightsheet, the expansion $\theta (\lambda)$
\begin{eqnarray}
\theta(\lambda)\equiv {d{\cal A}/d\lambda \over {\cal A}}
\end{eqnarray} %({\bf insert eqn (5.3) from Bousso})
can be allowed to be slightly positive for a short interval, as
long as the coarse grained $\theta (\lambda)$ is not positive.
(Alternatively, $\theta (\lambda)$ could be allowed to become
slightly positive, but with magnitude smaller than a coarse
graining scale.) This definition does {\it not} admit lightsheets
with diverging area (as displayed in Fig.~{\ref{fig2}}), and it
reduces to the usual covariant bound in simple cases. However,
using this coarse grained definition, the light sheets bound to
the brane do not terminate due to focusing. If a coarse graining
scale of order $d$ is adopted, they produce a uniform sheet of
thickness $2d$ (See Fig.~{\ref{fig4}}.) According to any coarse
grained definition, the $L(B)$ in the brane world construction of
\cite{HH} (which has the same geometry as the brane itself) is the
smooth limit of a family of lightsheets $L(B')$ resulting from
surfaces $B'$ with decreasing thickness in the extra dimension.

%\bigskip

%\noindent {\bf Review Covariant Bound and our application to brane
%worlds. Discuss focusing of null geodesics. Derive focusing
%theorem and propose coarse grained definition of lightsheets.}

%\noindent {\bf Need 3 figures: (1) the covariant bound geometry
%(2) focusing of light rays and divergent expansion, (3) binding of
%rays to brane} \vfill \eject

\begin{figure}
\epsfxsize=6in \centerline{\epsffile{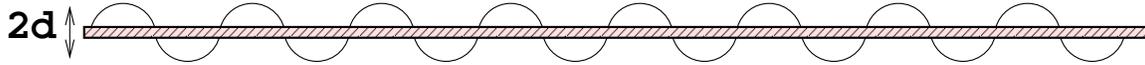}}
%
%\centerline{\includegraphics[scale=0.5]{hat.eps}}%
\caption{Light rays bound to the brane.}
 \label{fig4}
\end{figure}

%\eject

\bigskip

%%%%%%%%%%%%%%%%%%%%%%%%%%%%%%%%%%%%%%%%%%%%%%%%%%%%%%%%%%%%%%%%%
%%%
%%%                   ACKNOWLEDGEMENTS
%%%
%%%%%%%%%%%%%%%%%%%%%%%%%%%%%%%%%%%%%%%%%%%%%%%%%%%%%%%%%%%%%%%%%
\section*{Acknowledgements}
\noindent We thank D. Marolf for useful comments. The work of D.K.H. is supported by
KRF PBRG 2002-070-C00022. The work of S.H. was supported in part
under DOE contract DE-FG06-85ER40224.

%\newpage

\end{document}